\documentclass[epj]{webofc}
\usepackage[varg]{txfonts}   % Web of Conferences font
\woctitle{MESON2016 - the 14$^\textrm{th}$ International Workshop on Meson Production, Properties and Interaction}
\begin{document}
\selectlanguage{english}
\title{Inclusive production of J/$\psi$ and $\psi'$ mesons at the LHC}

% insert email only for speaker/presenter
\author{Anna Cisek\inst{1}\fnsep\thanks{\email{acisek@ur.edu.pl}} \and
        Antoni Szczurek\inst{1,2} 
}

\institute{University of Rzesz\'ow, PL-35-959 Rzesz\'ow, Poland
\and
           Institute of Nuclear Physics PAN, PL-31-342 Cracow, Poland 
          }

\abstract{%
We discuss the prompt production of $J/\psi$ mesons in proton-proton
collisions at the LHC within a NRQCD $k_{t}$-factorization approach
using Kimber-Martin-Ryskin (KMR) unintegrated gluon distributions (UGDF). 
We include both direct color-singlet production ($gg \to J/\psi$g) as
well as a feed-down from $\chi_{c} \to J/\psi \gamma$ and $\psi' \to J/\psi X$.
The production of the decaying mesons ($\chi_{c}$ or $\psi'$)is also
calculated within NRQCD $k_{t}$-factorization.
The corresponding matrix elements for $gg \to J/\psi$, $gg \to \psi'$ and $gg \to \chi_{c}$ 
include parameters of the nonrelativistic spatial wave functions of
quarkonia at $r=0$, which are taken from potential models from the literature.
We get the ratio of the corresponding of the cross sections for
$\chi_{c}(2)$-to-$\chi_{c}(1)$ much closer to experimental data than
obtained in recent analyses. Differential distributions in rapidity of $J/\psi$ and $\psi'$
are calculated and compared to experimental data of the ALICE and LHCb
collaborations. We discuss possible onset of gluon saturation effects at
forward/backward rapidities. One can describe the experimental data for $J/\psi$ production within
model uncertainties with color-singlet component only. Therefore our theoretical results leave only a relatively small room
for the color-octet contributions.
}
\maketitle
\section{Introduction}

For a long time there are discrepancies among authors about the production mechanism 
of $J/\psi$ quarkonia in proton-proton
and proton-antiproton collisions. Some authors think that the cross section is dominated by the 
color-octet contribution.
Some authors believe that the color-singlet contribution dominates. 
The color-octet contribution cannot be calculated
from first principle and is rather fitted to the experimental data. 
Different fits from the literature give different
magnitudes of the color-octet contributions. Therefore we concentrate on the color-singlet contribution.
In the present paper we wish to calculate the color-singlet contribution as well as 
possible in the NRQCD $k_t$-factorization and see how much room is left for the more 
difficult color-octet contribution.
In the present approach we concentrate rather on small transverse momenta of $J/\psi$ or $\psi'$ relevant for
ALICE and LHCb data \cite{Alice_2760, LHCb_7000, Alice_7000_a, Alice_7000_b, LHCb_13000}. We expect that color-singlet
contributions may dominate in this region of the phase space.
Finally $\psi'$ quarkonium also has a sizable branching fraction into $J/\psi X$ \cite{PDG}. Fortunately this contribution
is much smaller than the direct one as will be discussed in \cite{CS2016}. It was considered recently in an almost identical
approach in \cite{BLZ2015a}.

\section{Inclusive production of $J/\psi$ and $\psi'$ mesons in the NRQCD $k_t$-factorization approach}

\begin{figure}
\begin{center}
\includegraphics[width=4.75cm]{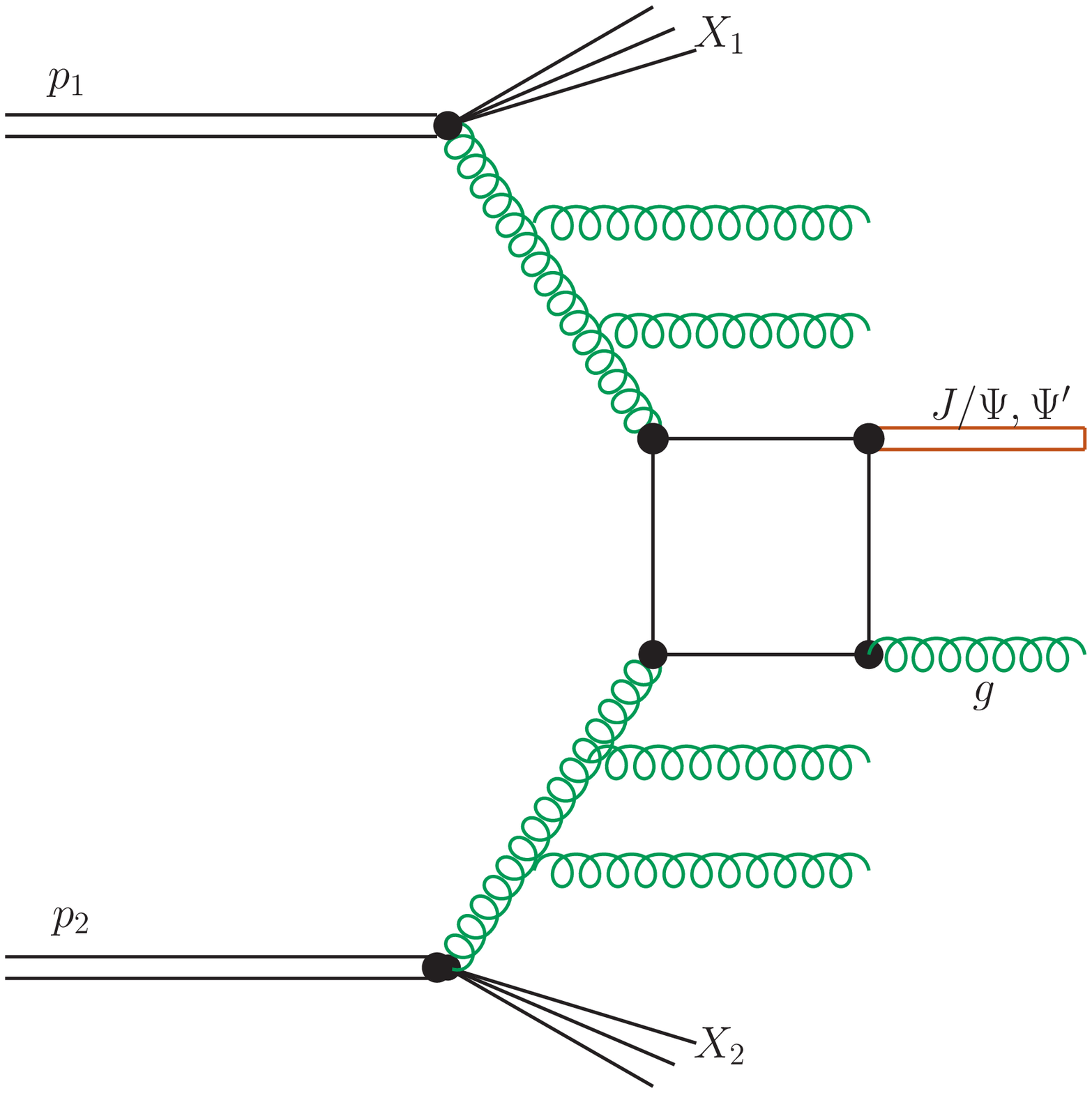} 
\includegraphics[width=5.5cm]{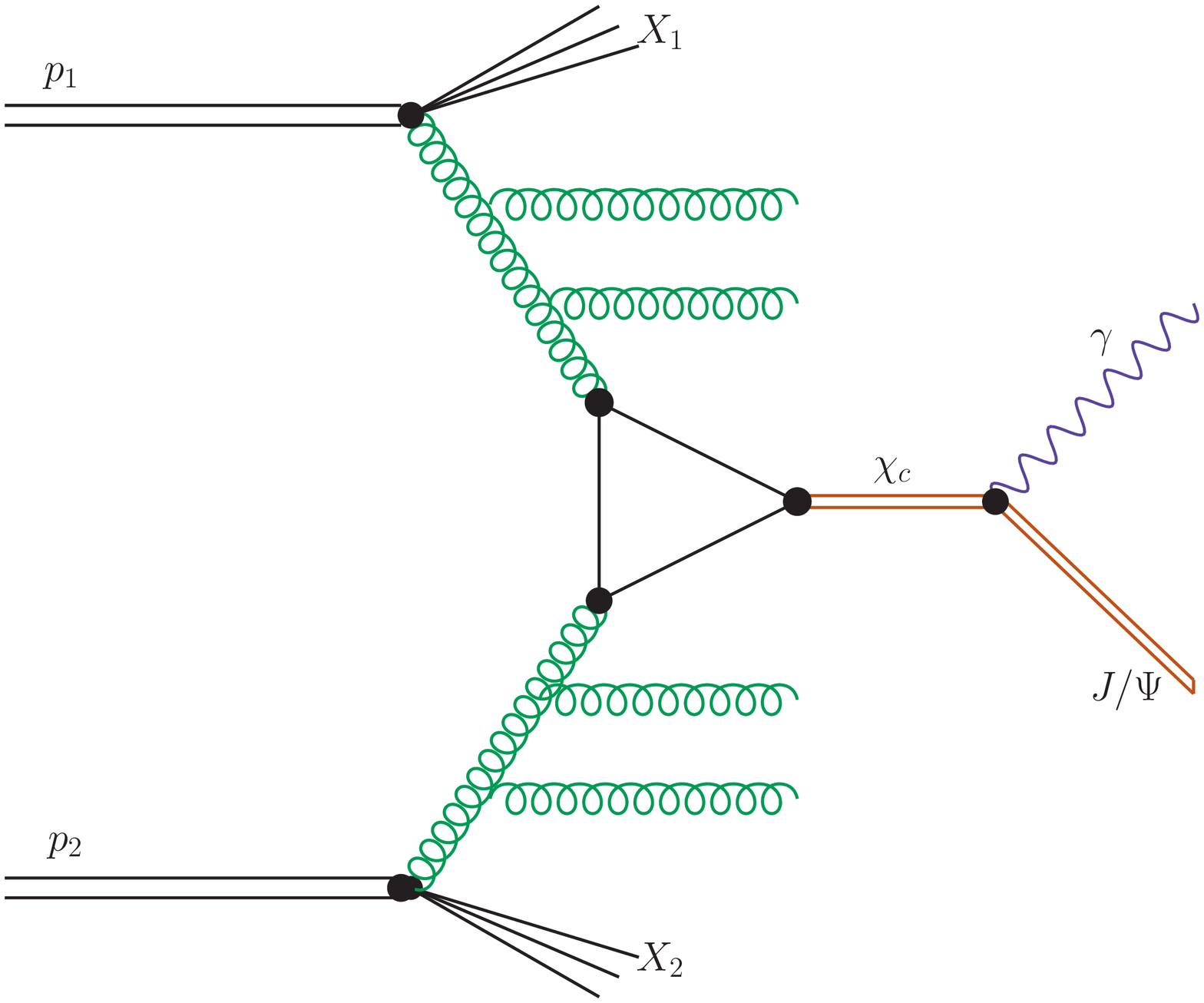} 
\end{center}
\caption{The leading-order diagram for prompt $J/\psi$ ($\psi'$) 
meson production in the $k_t$-factorization approach.}
\label{fig_diag1}
\end{figure}

The main color-singlet mechanism for the production of $J/\psi$ and $\psi'$ mesons is shown
in Fig.\ref{fig_diag1} (left panel). We restrict ourselves to the gluon-gluon fusion mechanism.
In the NLO the differential cross section in the $k_{t}$-factorization
can be written as:
\begin{eqnarray}
\frac{d \sigma(p p \to J/\psi g X)}{d y_{J/\psi} d y_g d^2 p_{J/\psi,t} d^2 p_{g,t}}
&& = 
\frac{1}{16 \pi^2 {\hat s}^2} \int \frac{d^2 q_{1t}}{\pi} \frac{d^2 q_{2t}}{\pi} 
\overline{|{\cal M}_{g^{*} g^{*} \rightarrow J/\psi g}^{off-shell}|^2} 
\nonumber \\
&& \times \;\; 
\delta^2 \left( \vec{q}_{1t} + \vec{q}_{2t} - \vec{p}_{H,t} - \vec{p}_{g,t} \right)
{\cal F}_g(x_1,q_{1t}^2,\mu_{F}^{2}) {\cal F}_g(x_2,q_{2t}^2,\mu_{F}^{2}) \; .
\label{kt_fact_gg_jpsig}
\end{eqnarray}
We calculate the dominant color-single $g g \to J/\psi g$ contribution taking 
into account transverse momenta of initial gluons.
The corresponding matrix element squared for the $g g \to J/\psi g$ is
\begin{equation}
|{\cal M}_{gg \to J/\psi g}|^2 \propto \alpha_s^3 |R(0)|^2 \; .
\label{matrix_element} 
\end{equation}
The matrix element is taken from \cite{Baranov_2002}.
In our calculation we choose the scale of the running coupling constant as:
\begin{equation}
\alpha_s^3 \to \alpha_s(\mu_1^2) \alpha_s(\mu_2^2) \alpha_s(\mu_3^2) \; ,
\end{equation}
where $\mu_1^2 = max(q_{1t}^2,m_t^2)$,
      $\mu_2^2 = max(q_{2t}^2,m_t^2)$ and
      $\mu_3^2 = m_t^2$,
where here $m_t$ is the $J/\psi$ transverse mass.
The factorization scale in the calculation was taken as
$\mu_F^2 = (m_t^2 + p_{t,g}^2)/2$.

Similarly we calculate the P-wave $\chi_c$ meson production. Here the lowest-order subprocess
$g g \to \chi_c$ is allowed by positive $C$-parity of $\chi_c$ mesons.

In the $k_t$-factorization approach the leading-order cross section for 
the $\chi_c$ meson production can be written as:
\begin{eqnarray}
\sigma_{pp \to \chi_c} = \int d y d^2 p_t d^2 q_t \frac{1}{s x_1 x_2}
\frac{1}{m_{t,\chi_c}^2}
\overline{|{\cal M}_{g^*g^* \to \chi_c}|^2} 
{\cal F}_g(x_1,q_{1t}^2,\mu_F^2) {\cal F}_g(x_2,q_{2t}^2,\mu_F^2) / 4
\; ,
\label{useful_formula}
\end{eqnarray}
which can also be used to calculate rapidity and transverse momentum distributions 
of the $\chi_c$ mesons.
In the last equation ${\cal F}_g$ are unintegrated gluon distributions and 
$\sigma_{g g \to \chi_c}$ is $g g \to \chi_c$ (off-shell) cross section.
The situation is illustrated diagrammatically in Fig.\ref{fig_diag1} (right panel).

The matrix element squared for the $g g \to \chi_c$ subprocess is
\begin{equation}
|{\cal M}_{gg \to \chi_c}|^2 \propto \alpha_s^2 |R'(0)|^2 \; .
\label{matrix_element} 
\end{equation}
We used the matrix element taken from the Kniehl, Vasin and Saleev paper \cite{KVS2006}.

For this subprocess the best choice for running coupling constant is:
\begin{equation}
\alpha_s^2 \to \alpha_s(\mu_1^2) \alpha_s(\mu_2^2) \; ,
\end{equation}
where $\mu_1^2 = max(q_{1t}^2,m_t^2)$ and
      $\mu_2^2 = max(q_{2t}^2,m_t^2)$.
Above $m_t$ is transverse mass of the $\chi_c$ meson.

The factorization scale for the $\chi_c$ meson production is fixed as $\mu_F^2 = m_t^2$.

\section{Results}

\begin{figure}
\begin{center}
\includegraphics[width=4.5cm]{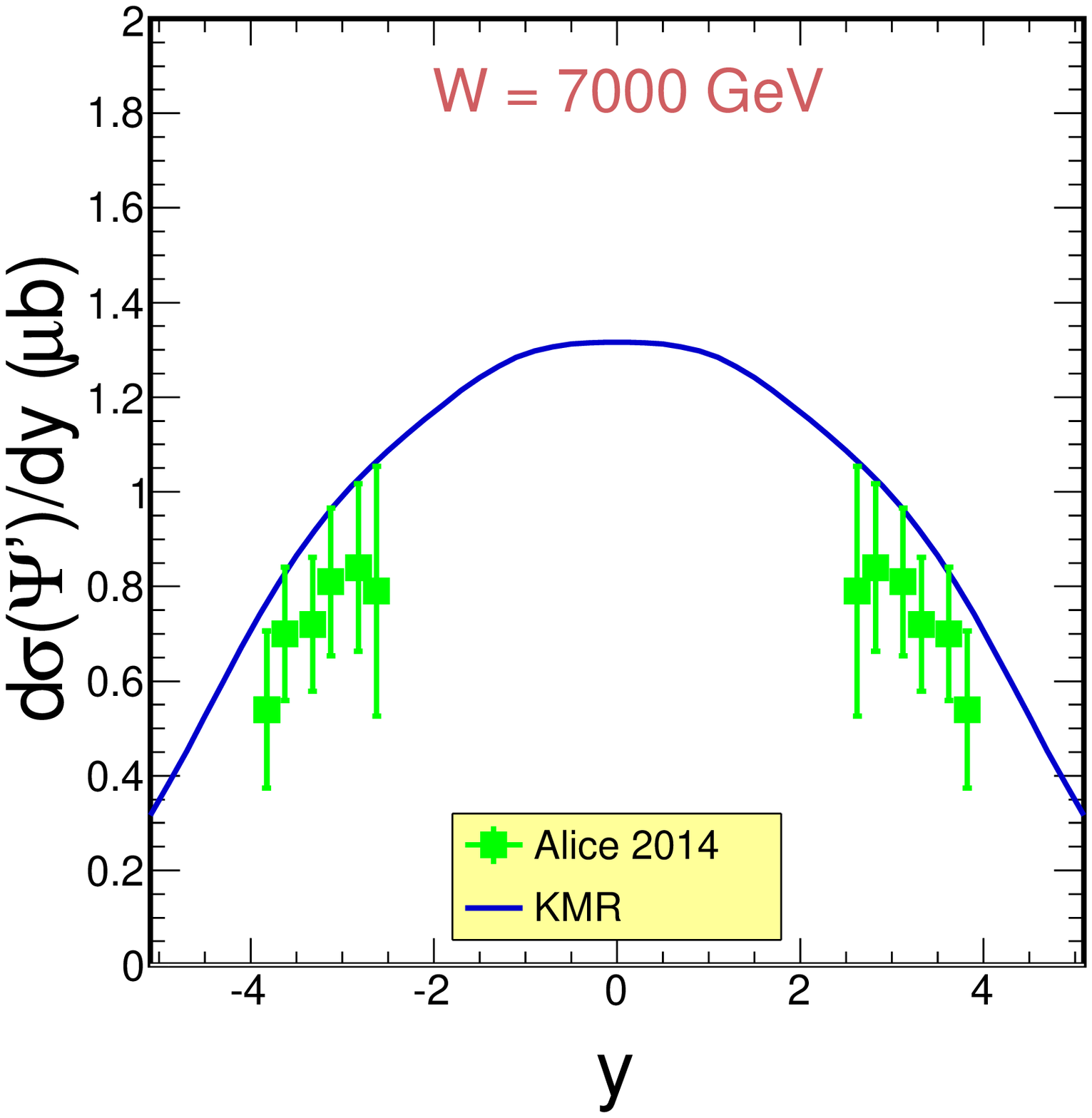} 
\includegraphics[width=4.5cm]{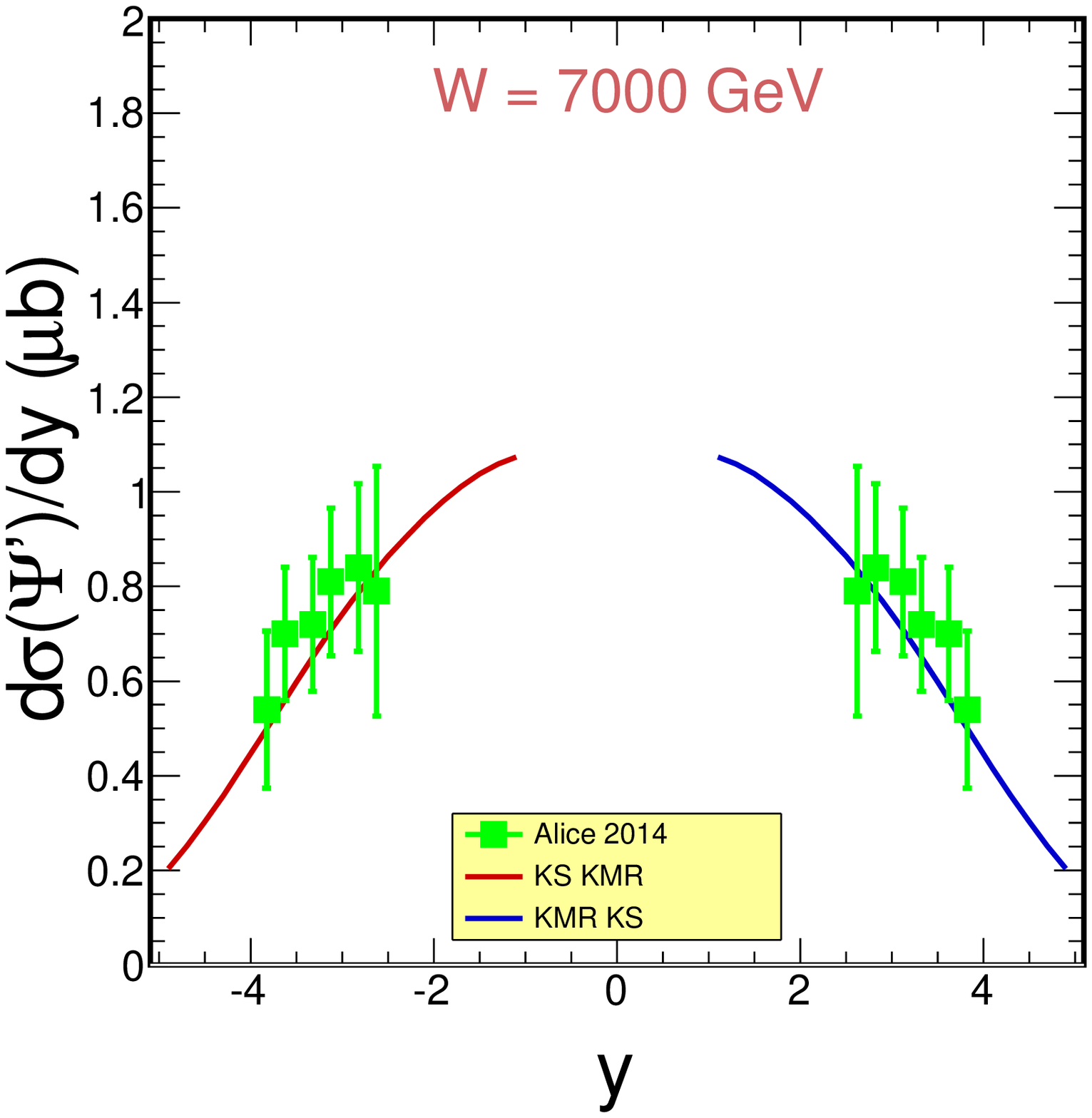}
\end{center}
\caption{Rapidity distribution of $\psi'$ meson with KMR (left plots) and mixed UGDFs
(KS and KMR, right plots). The ALICE data \cite{Alice_7000_b} are shown for comparison.}
\label{fig_dsig_dy_psi2S}
\end{figure}

In Fig.\ref{fig_dsig_dy_psi2S} we show differential cross section in
rapidity for $\psi'$ production at 7 TeV.
Our results are compared with ALICE experimental data \cite{Alice_7000_b}. In the left panel we present results
for Kimber-Martin-Ryskin (KMR) UGDF and in the right panel for mixed Kimber-Martin-Ryskin (KMR) and Kutak-Stasto (KS) UGDFs.
Because KMR alone overshoot experimental data for rapidity distribution the best solution is to take the KMR
distribution for large x and KS for small x. For $\psi'$ meson we have to include only the direct diagram
so it's easy to compare our result with experimental data.

For $J/\psi$ meson we have to include both diagrams. Below we present
results for these two subprocesses.
\begin{figure}
\begin{center}
\includegraphics[width=4.5cm]{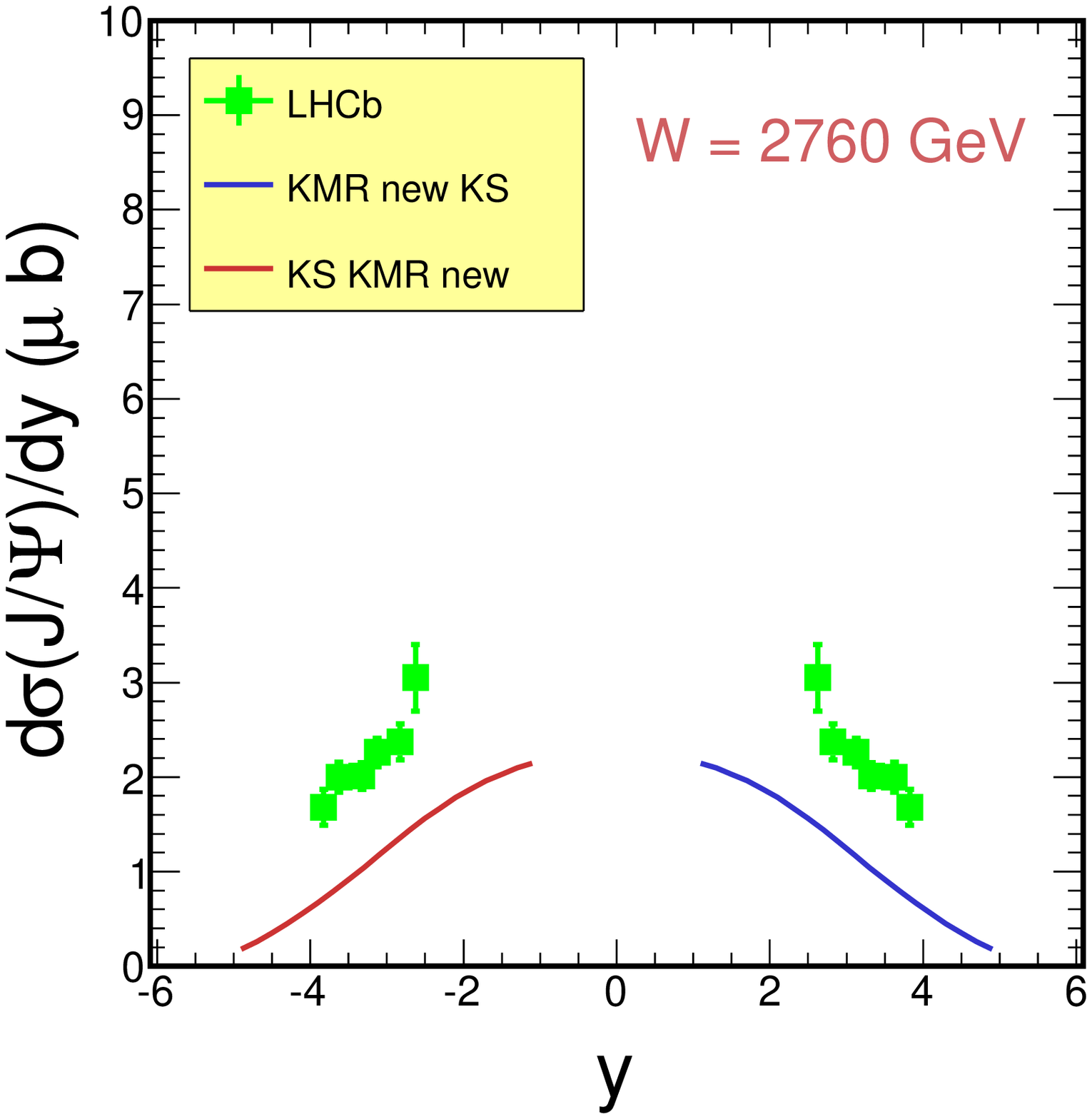} 
\includegraphics[width=4.5cm]{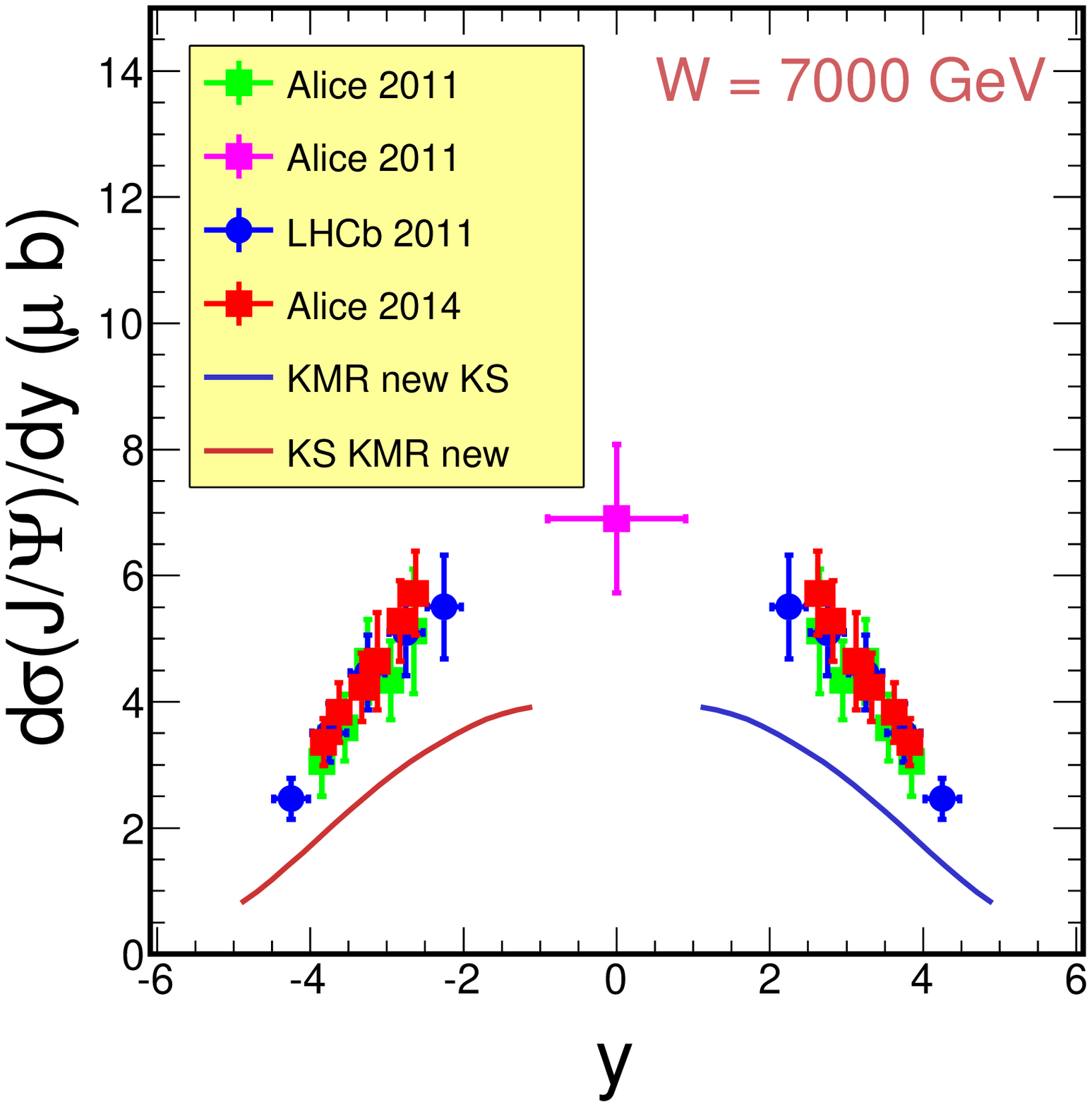} 
\includegraphics[width=4.5cm]{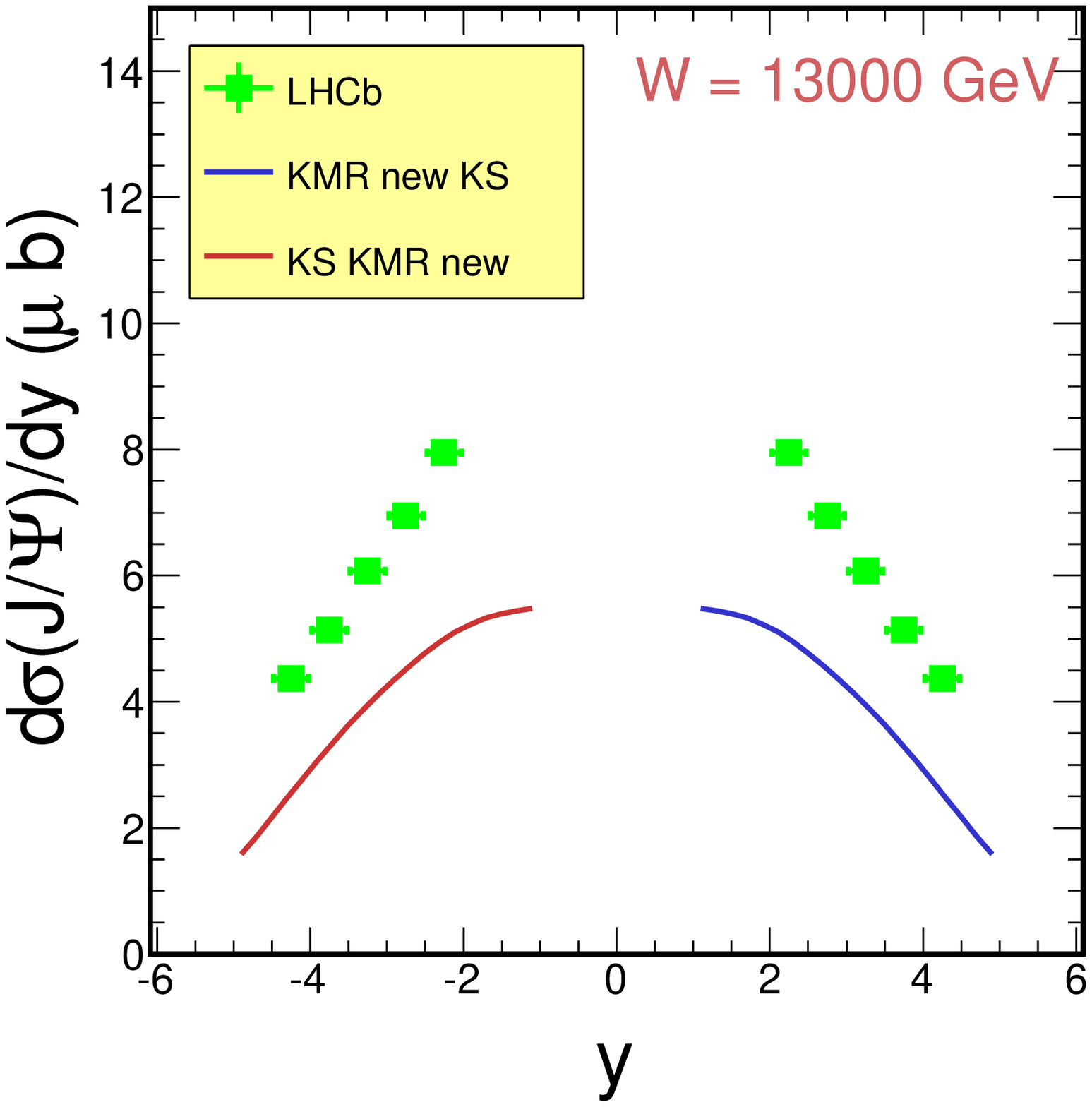} 
\end{center}
\caption{Rapidity distribution of $J/\psi$ meson with KMR (upper plots) 
and mixed UGDFs (Kutak-Stasto and KMR).
The ALICE and LHCb data points \cite{Alice_2760, LHCb_7000, Alice_7000_a, Alice_7000_b, LHCb_13000}
are shown for comparison.}
\label{fig_dsig_dy_jpsi}
\end{figure}
In Fig.\ref{fig_dsig_dy_jpsi} we show rapidity distribution for direct $J/\psi$ meson production.
We present results for three different values of energy: W = 2.76 TeV (left), W = 7 TeV (middle)
and W = 13 TeV (right). Our results are compared with ALICE and LHCb experimental data
\cite{Alice_2760, LHCb_7000, Alice_7000_a, Alice_7000_b, LHCb_13000}.

\begin{figure}
\begin{center}
\includegraphics[width=4.5cm]{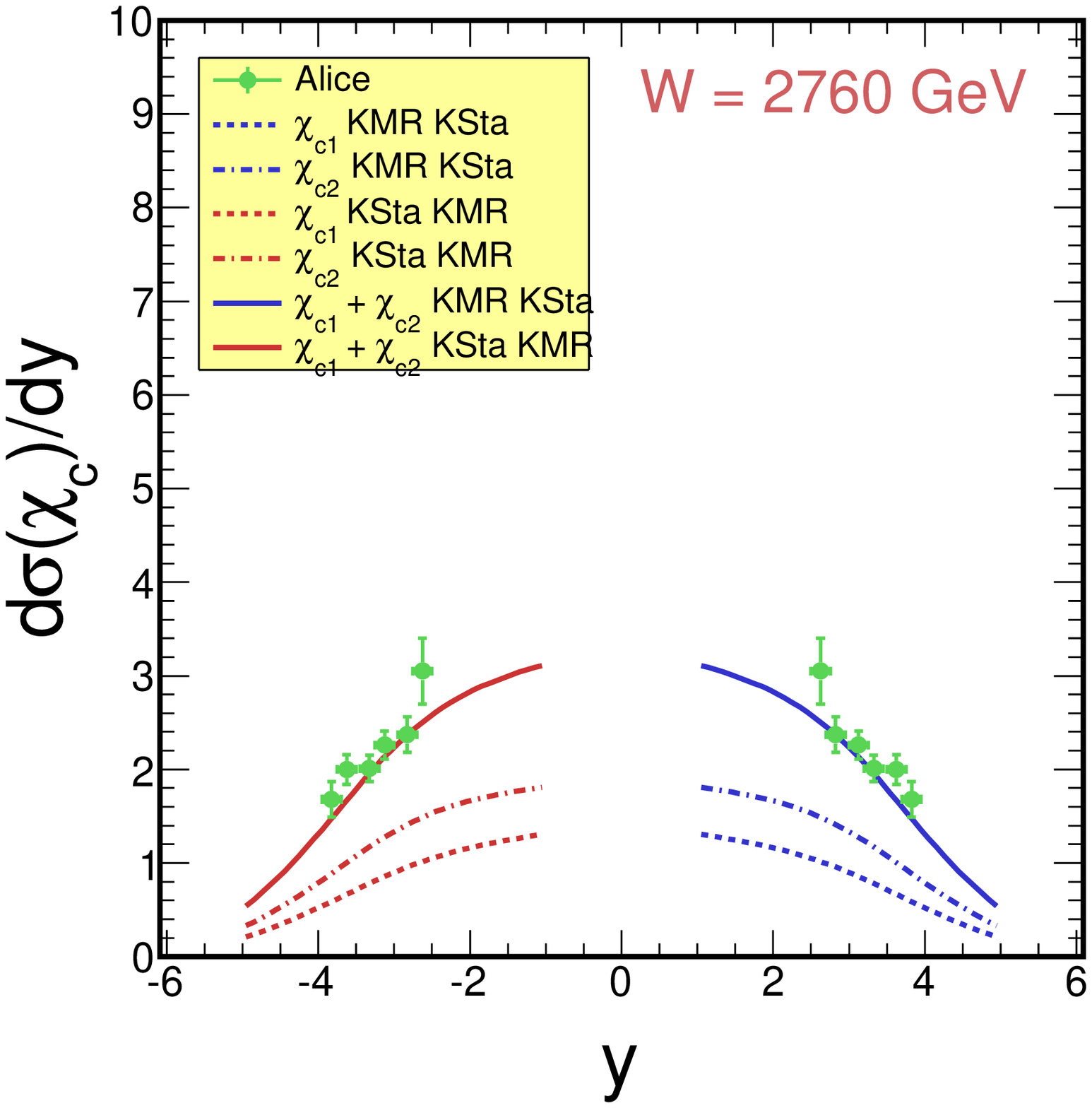} 
\includegraphics[width=4.5cm]{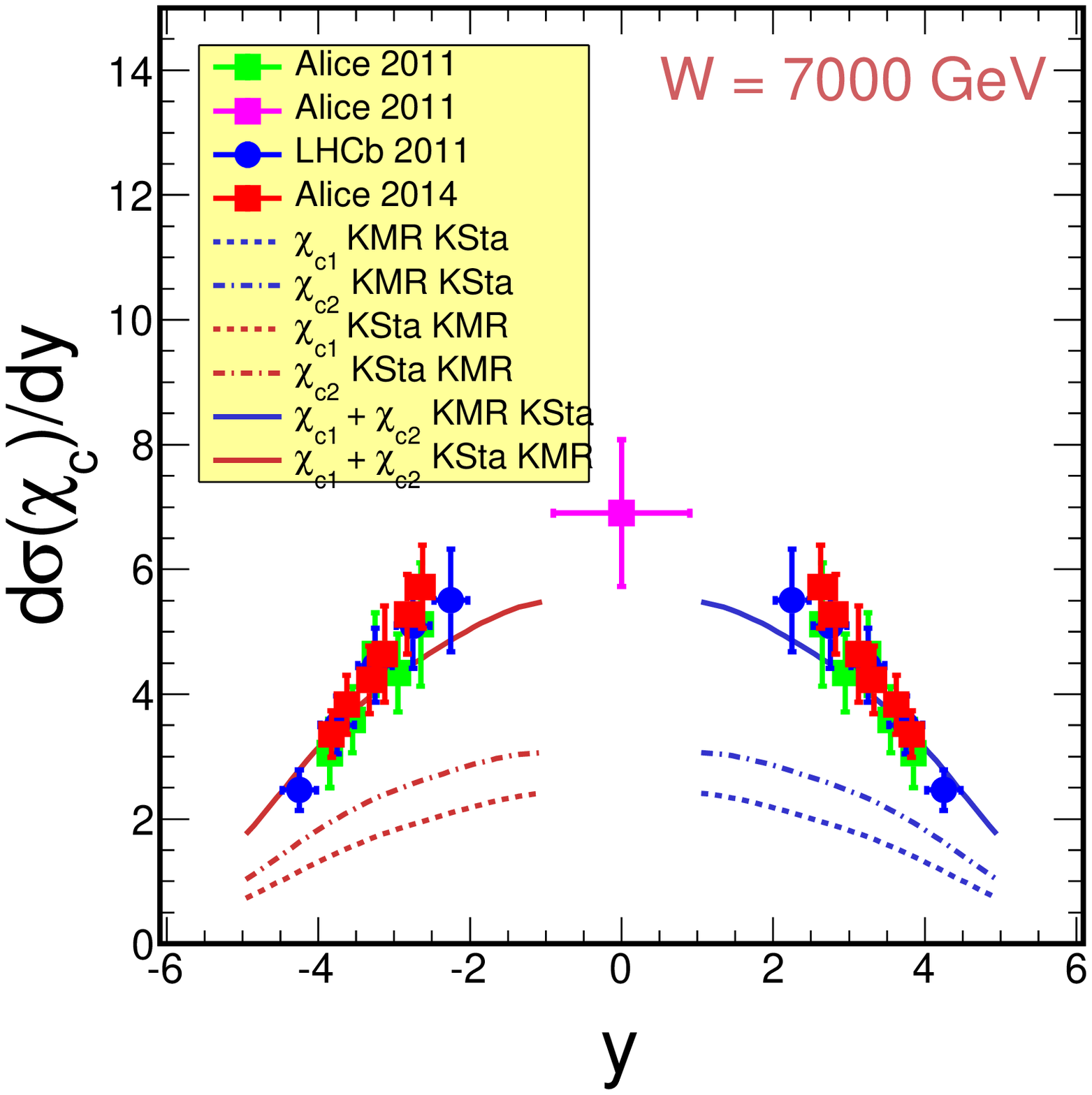} 
\includegraphics[width=4.5cm]{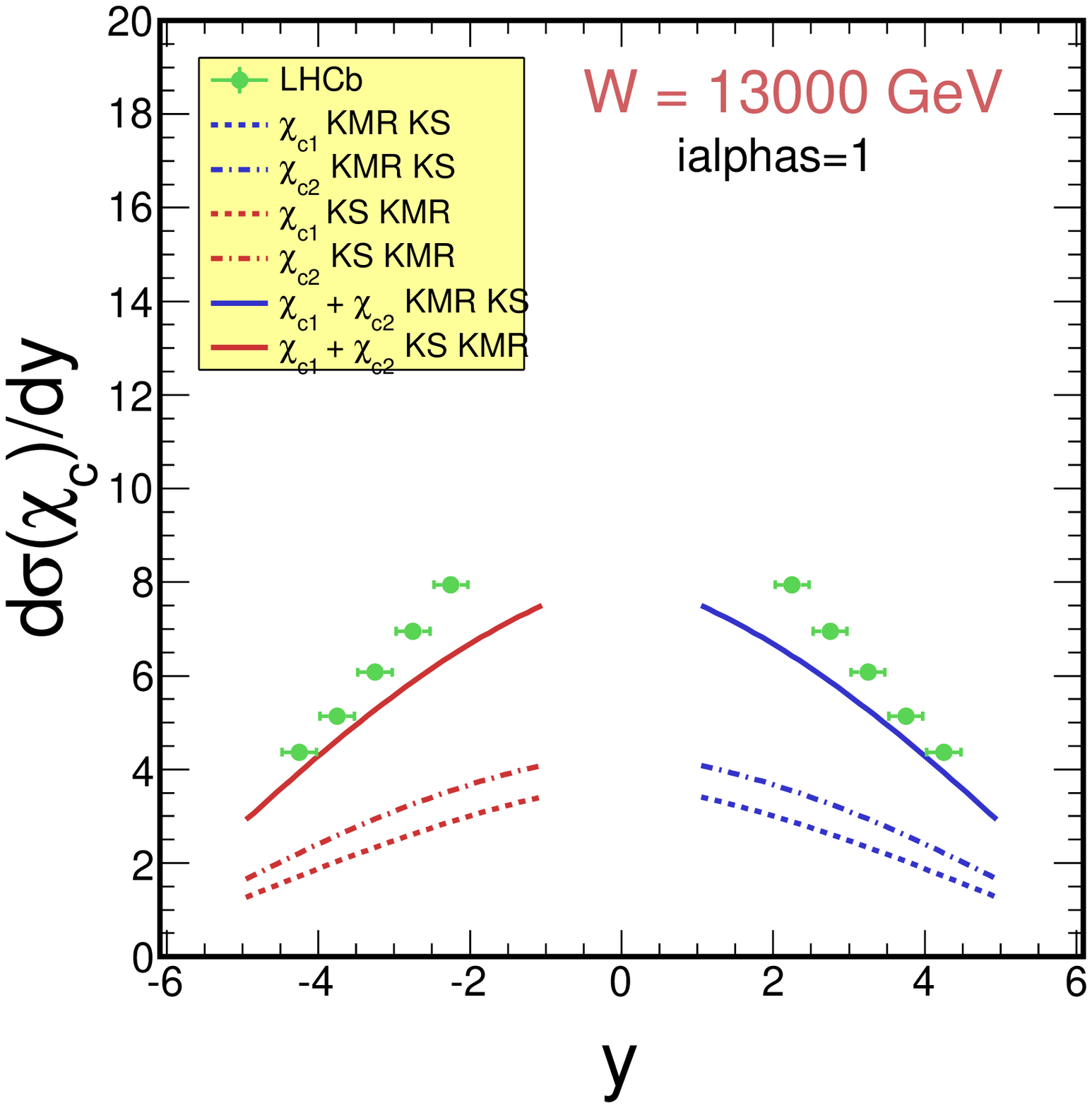} 
\end{center}
\caption{Rapidity distribution of $\chi_{c}$ meson with mixed UGDFs (Kutak-Stasto and KMR).
The ALICE and LHCb data points \cite{Alice_2760, LHCb_7000, Alice_7000_a, Alice_7000_b, LHCb_13000}
for $J/\psi$ are shown for comparison.}
\label{fig_dsig_dy_chic}
\end{figure}

In Fig.\ref{fig_dsig_dy_chic} we present results for three different values of energy:
W = 2.76 TeV (left), W = 7 TeV (middle) and W = 13 TeV (right) panel. The dotted lines are for $\chi_{c1}$
meson contribution, the dot-dashed lines are for $\chi_{c2}$ meson contributions ant the solid lines are sum of these two
components. The presented here results are calculated with mixed UGDFs (KMR and KS).

\section{Conclusion}

We have calculated the color-singlet contribution in the NRQCD $k_t$-factorization and compared our results
with ALICE and LHCb data. Our results in rapidity are almost consistent
or even exceed  experimental data.
Cross section strongly depends on UGDF and we think the best solution is to use mixed UGDFs (KMR-KS).
In our approach only small room is left for color-octet contribution.

\end{document}